\documentclass[11pt]{article}  

\usepackage[T1]{fontenc}

\usepackage{graphicx}
\usepackage{mathtools}
\usepackage{psfrag}
\usepackage{subfig}

\usepackage{xspace}
\usepackage{pgfplots}
\usepackage{authblk}

\usepackage{enumitem}
\setitemize{label=\scriptsize{$\blacksquare$},topsep=0em}
\setenumerate{label=(\alph*), topsep=0em}

\newcommand{\R}{\mathbb R}

\newcommand{\rmd}{\mathrm d}

\newcommand{\supp}{\mbox{supp}}

\usepackage{soul}
\usepackage{xcolor}
\colorlet{lred}{red!40}
\colorlet{lgreen}{green!40}
\colorlet{lblue}{blue!40}

\definecolor{lime}{RGB}{0,142,192}
\definecolor{goyel}{RGB}{229,211,80}
\definecolor{blor}{RGB}{250,188,81}
\definecolor{blgr}{RGB}{83,154,85}
\definecolor{orred}{RGB}{249,77,25}
\definecolor{orbl}{RGB}{238,118,0}
\definecolor{dot}{RGB}{0,0,95}

\def\mirrord at (#1,#2){\filldraw[color=orred](#1-0.07,#2)--(#1-0.07,#2-0.2)--(#1-0.12,#2-0.2)--(#1,#2-0.35)--(#1+0.12,#2-0.2)--(#1+0.07,#2-0.2)--(#1+0.07,#2)--cycle;}

\newcommand{\equic}[3][1 cm]{
  \foreach \i in {1,...,#2} {
    \coordinate (N\i) at ({#1*cos(\i*180/(#2+1))+#3cm},{-#1*sin(\i*180/(#2+1))});
    \fill[black] (N\i) circle (0.1 cm);
}
}

\newcommand{\equicfull}[3][1 cm]{
  \foreach \i in {1,...,#2} {
    \coordinate (N\i) at ({#1*cos(\i*360/(#2))+#3cm},{-#1*sin(\i*360/(#2))});
    \fill[dot] (N\i) circle (0.1 cm);
}
}

\usepackage{amsmath,amsfonts,amssymb}
\usepackage{graphicx}
\usepackage[colorlinks=true, allcolors=blue]{hyperref}

\title{Learned backprojection for sparse and limited view photoacoustic tomography }

\author{Johannes Schwab, Stephan Antholzer, Markus Haltmeier}

\affil{Department of Mathematics, University of Innsbruck\\
Technikerstra{\ss}e 13, 6020 Innsbruck, Austria, {\tt johannes.schwab@uibk.ac.at}}
\date{}

\begin{document}
 
\maketitle

\begin{abstract}
Filtered backprojection (FBP) is an efficient and popular class of tomographic image reconstruction methods. In photoacoustic tomography, these algorithms are based on theoretically exact analytic inversion formulas which results in accurate reconstructions. However, photoacoustic measurement data are often incomplete (limited detection view and sparse sampling), which results in artefacts in the images reconstructed with FBP. In addition to that, properties such as directivity of the acoustic detectors are not accounted for in standard FBP, which affects the reconstruction quality, too. To account for these issues, in this papers we propose to improve FBP algorithms based on machine learning techniques. In the proposed method, we include additional weight factors in the FBP, that are optimized on a set of incomplete data and the corresponding ground truth photoacoustic source. Numerical tests show that the learned FBP improves the reconstruction quality compared to the standard FBP.
\end{abstract}


\section{Introduction}
\label{sec:intro}  

Photoacoustic tomography (PAT) is a promising imaging method for medical diagnosis based on the photoacoustic effect. The process of data acquisition for PAT can be described as follows. Illumination of the sample with a short laser pulse leads to heating of the tissue, and along with that to a thermoelastic expansion. This expansion generates a pressure wave, which is measured outside of the tissue.
In the case of full data, several efficient reconstruction methods have been developed. In particular, this includes   filtered backprojection algorithms bases on analytic inversion formulas known for special geometries\cite{nguyen2009family,finch2007inversion,finch2004determining,haltmeier2014universal,kunyansky2007explicit,xu2005universal}.\\
In practice it is often not possible to collect measurement data of the resulting pressure wave on a measurement surface surrounding the whole object, and the data is known only on parts of such a surrounding surface. This problem of image reconstruction from incomplete data of this type, is called limited view problem. Furthermore, because for each spatial measurement an individual, often costly, detector is needed, the number of detectors placed on the detection area is limited, leading to spacially sparsly sampled data. In particular, in the reconstruction of the initial pressure distribution, the application of existing FBP algorithms to the limited view problem as well as the sparse data problem leads to specific artifacts. It is known, that the reconstruction of singularities not visible from the detector positions is unstable, whereas singularities contained within the convex hull can theoretically be stably recovered in appropriate spaces \cite{kuchment2011mathematics}.\\
For PAT from limited view data several iterative reconstruction methods have been proposed to reduce limited view artifacts and improve the reconstruction quality \cite{paltauf2007experimental,dean2012accurate, hauptmann2018model}. 
In \cite{paltauf2009weight} the authors proposed using weight factors depending on the angle between the reconstruction- and detection point for an inversion formula, which is exact on continuously sampled data given on the whole boundary. This method tries to compensate the zero-extension of the data on parts of the boundary, where the data is not available. In \cite{liu2013limited} an iterative algorithm, updating these angle-depending weight factors has been proposed. Deep learning methods for artifact reduction in limited view PAT have been proposed in \cite{schwab2018dalnet} and a framework to learn an extension of the data was introduced in \cite{dreier2017operator}. In this paper we propose to learn weight factors from a big dataset of incomplete data with simulated directional detector sensitivity and the corresponding ground truth reconstructions. We compute appropriate weights via a gradient descent method in the framework of neural networks and the available software and implementations on GPUs.

\section{Methods}

\subsection{Photoacoustic tomography (PAT)}

In PAT the investigated sample is illuminated with a short laser pulse, which induces heating in the tissue which in turn causes a thermoelastic expansion. This expansion generates a pressure wave, which satisfies the initial value problem

\begin{align}
\partial_t^2 u(x,t)-c^2\Delta_x u(x,t)&=0,\nonumber\\
u(x,0)&=f(x),\label{eq:waveeq}\\
\partial_t u(x,0)&=0.\nonumber
\end{align}

\noindent Here for $d\in\{2,3\}$ the spatial location is denoted by $x\in\R^d$, $t\in\R$ denotes the time, $\Delta_x$ the spatial Laplacian and $c$ the speed of sound. We further assume, that the photoacoustic (PA) source satisfies $\supp{f}\subset\Omega$ for a bounded domain $\Omega \subset \R^d$. \\

Spatial dimension $d=3$ is relevant for PAT, when point-wise measurements are taken on the detection surface $\partial \Omega$. In this paper we consider two spatial dimension, which is relevant for PAT using integrating line detectors \cite{bauer2017all, burgholzer2007temporal, paltauf2007photoacoustic, paltauf2017piezoelectric}. In the case of integrating line detectors, the PA source $f$ corresponds to a projection image of the 3D source. By reconstructing projection images of sufficiently many directions, the 3D source can be reconstructed using the inverse Radon transform.\\

In two spatial dimensions the solution of \eqref{eq:waveeq} denoted by $u_f$ for given initial pressure $f$ is given by the well known solution formula
\begin{equation}\label{eq:solform}
u_f(x,t)\coloneqq \frac{1}{2\pi} \frac{\partial}{\partial t}\int_0^t\int_{\partial B_1(0)}\frac{rf(x+r\omega)}{\sqrt{t^2-r^2}} \rmd \omega \ \rmd r,
\end{equation}
where $\partial B_1(0)$ denotes the unit sphere. From data as in \eqref{eq:solform},  the PA source can be reconstructed by exact inversion formulas. 
In this paper, we consider the following three issues affecting  PAT image reconstruction.  We denote points on detection curve by $s \in \partial \Omega$.\\

\begin{itemize}
\item \textbf{Limited view PAT}\\
For a measurement surface $\partial \Omega$ containing the support of the source $f$, the problem of recovering $f$ from measurements on parts of $\partial \Omega$ is called limited view PAT. For limited view PAT, the given data is of the form
\begin{align}
u_f(s_i,t_j) \quad \quad i=1,\ldots, N_s; \ j=1,\ldots N_t \text{  with  } s_i\in\Gamma,
\end{align} 
where the detector locations for $i=1,\ldots N_s$ are contained in a subset $\Gamma\subset\partial \Omega$, which is not surrounding the source $f$. (see Figure \ref{fig:meas} \textbf{A)})

\item \textbf{Sparse data PAT}\\
The number of spatial measurements of the solution of \eqref{eq:waveeq} on the measurement curve $\partial \Omega$ is limited, which is called sparse data problem. The data for the reconstruction of a acoustic source $f$ is therefore given by
\begin{align}
\{u_f(s_i,t_j)\mid i=1,\ldots, N_s; j=1,\ldots N_t\} \text{  with  } s_i\in\Gamma,
\end{align} 
where the number of detectors $N_s$ is small, whereas the number of measurements in time $N_t$ is typically large enough. (see Figure \ref{fig:meas} \textbf{B)})

\item \textbf{Detector directivity}\\
To model the directional sensitivity of the acoustic detectors we consider a weight function depending on the angle between the detection points $s$ and source locations. In particular we assume, that the wavedata is given by \cite{xu2004time,zangerl2018photoacoustic}
\begin{equation}\label{eq:dirsol}
u_f(s,t)\coloneqq \frac{1}{2\pi} \frac{\partial}{\partial t}\int_0^t\int_{\partial B_1(0)}\frac{rf(s+r\omega)}{\sqrt{t^2-r^2}} \varphi( \nu_s,\omega) \ \rmd \omega \ \rmd r,
\end{equation}
where $\varphi:\R^2\times \R^2 \rightarrow[0,1]$ is a function depending on the angle between source location and the outer normal on the detection surface at $s$. In our simulations with a circle as detection curve, we modelled the sensitivity by choosing the function $\varphi$ as
\begin{align}
\varphi(s,\omega)\coloneqq
\begin{cases} 
\frac{\langle s-\omega,\omega\rangle^2}{\|s-\omega\|^2}=\cos(\alpha)^2, \quad &|\alpha|<\frac{\pi}{2},\\
0,  & |\alpha|\geq\frac{\pi}{2},
\end{cases}
\end{align}
where $\alpha$ denotes the angle between source and detection point. 

\begin{figure}[h]
\begin{minipage}[t]{0.45\linewidth}
\begin{tikzpicture}[scale=0.45]
\draw (-8,0) node[left]{};
\draw [line width=1.2pt](0,0) circle(4.2);
\filldraw[color=black] (0,0) circle(0.1); 
\filldraw[color=black] (-4.2/2,-1.73205080757*4.2/2) circle(0.1);
\draw (4.2,0) node[right]{\large $\partial \Omega$};
\draw (-4.4/2,-1.8*4.2/2) node[below]{$s$};
\draw [color= gray] (-4.2/2,-1.73205080757*4.2/2) circle(3.5);
\draw (0,0)--(-4.2/2,-1.73205080757*2.1);
\filldraw[color=black] (-2.1+3.5*0.9781476,-1.73205080757*2.1+3.5*0.20791169) circle(0.1);
\draw (-4.2/2,-1.73205080757*2.1)--(-2.1+3.5*0.9781476,-1.73205080757*2.1+3.5*0.20791169);
\draw (-2.1+3.5*0.9781476,-1.7*2.1+3.5*0.21) node[right] {$r\omega$};
\draw (-0.3,-2.85) node[above] {$\alpha$};
\end{tikzpicture}
\end{minipage}
\hfill
\begin{minipage}[t]{0.45\linewidth}
\begin{tikzpicture}
\draw[<->] (-3,-0)--(3,-0);
\draw[->] (0,-0.1)--(0,2.5);
\draw[domain=-{pi/2}:{pi/2},smooth,variable=\x] plot(\x,{2*cos(deg(\x))*cos(deg(\x))});
\draw (0,-0.1) node[below] {0};
\draw ({pi/2},-0.1) node[below]{$\pi/2$};
\draw ({-pi/2},-0.1) node[below]{$-\pi/2$};
\end{tikzpicture}
\end{minipage}
\caption{Illustration of the directivity model used for our simulations.}
\end{figure}
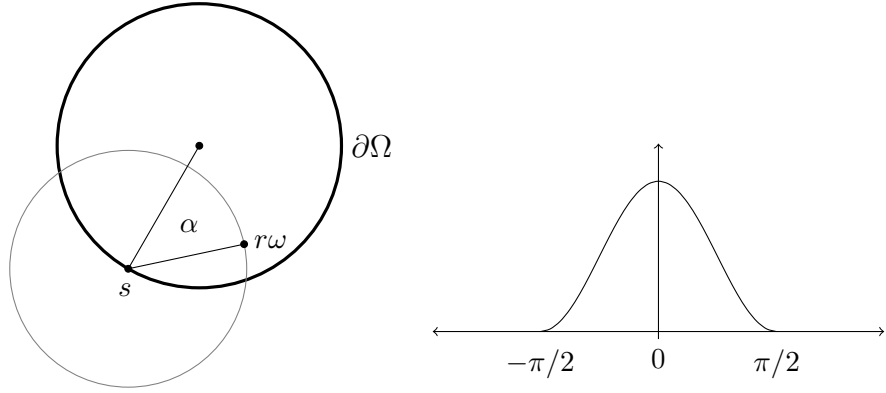

\end{itemize}

\subsection{Weighted universal backprojection (UBP)}

In the case, where the data of a source $f$ is given as a continuous function $g:\partial\Omega\times[0,\infty)\rightarrow \R$, defined on the whole boundary, the inverse problem of determining the source $f$ is uniquely and stably solvable. Many efficient and robust reconstruction methods for recovering the PA source including filtered backprojection, Fourier methods or time reversal have been developed \cite{}. In particular an exact inversion formula for continuous sampled data on certain detection surfaces $\partial\Omega$, including circular geometries, is the universal backprojection (UBP) fomula. In 2D the UBP has first been derived in \cite{burgholzer2007temporal}, and \cite{xu2005universal}, for the 3D case.\\

\noindent For wave data $g$, given on a subset of the boundary $\Gamma\subset\Omega$, the 2D universal backprojection (UBP) including a weight function $w: \R^2\times \partial{\Omega}\rightarrow \R$, is defined by 

\begin{align}
\Psi(w,g)(x)= -\frac{2}{\pi} \int_{\Gamma} w(x,s)^2 \langle \nu_s, x-s \rangle \int_{\|x-s\|}^\infty \frac{\partial_t(t^{-1} g(s,t))}{\sqrt{t^2-\|x-s\|^2}}\rmd t \ \rmd s.
\end{align}
Here $x$ denotes the reconstruction point, $\nu_s$ the exterior normal vector to a point $s\in\Gamma\subset\partial{\Omega}$ on the detection surface and $\langle \, \cdot \, , \, \cdot \, \rangle$ the inner product in $\R^2$. The weight function $w$ in the backprojection depends on the the reconstruction point and the detector point. In the standard UBP the weight function is constant, $w\equiv 1$. \\

\noindent In practice, for PA image reconstruction, the measurements of the pressure are only given for a finite number of detector locations, often not surrounding the whole object, and for a finite number of time samples. We denote by $G\in\R^{N_t\times N_s}$ the data for time samples $\{t_k=kT/N_t \mid k=1,\ldots, N_t\}\subset[0,T]$, with final measurement time $T\in\R$, and detector locations $\{s_k \mid k=1,\ldots, N_s\}\subset\Gamma$ on the measurement curve. The image reconstruction problem of PAT can now be modelled as estimating the discretized PA source $F\in\R^{N_x\times N_x}$ corresponding to the initial pressure $f$ in \eqref{eq:waveeq}, from discrete data
\begin{align}
G=A(F).
\end{align}
Here $A: \R^{N_x\times N_x} \rightarrow \R^{N_t\times N_s}$ the discretized operator mapping a source to the corresponding discrete wave data. Discetizing the weighted UBP formula $\Psi$ leads to a mapping $P(\mathbb{W},\cdot): \R^{N_t\times N_s}\rightarrow \R^{N_x\times N_x}$ with a weight vector $\mathbb{W}\in \R^{N_x\times N_x \times N_s}$. The case $\mathbb{W}\equiv 1$ corresponds to the discretization of the ordinary UBP.

\subsection{Learning the weights in the UBP}

Deep learning and in particular deep convolutional neural networks (CNNs) have been very successfully applied to a great variety of pattern recognition and image processing tasks. Recently a lot of research has been done in solving inverse problems, incorporating deep learning techniques, including efficient and accurate image reconstruction methods in tomographic problems \cite{antholzer2018deep, chen2017low, han2017deep, hauptmann2018model, kelly2017deep, zhang2016image,li2018nett}.\\
In the context of a FBP in limited data, the task of learning is to find optimal weights in the discretized UBP formula, with respect to a data set $(G_i,F_i)_{i=1}^M\subset \R^{N_s\times N_t}\times\R^{N_x\times N_x}$, where $G_i, i=1,\ldots, M$ denote the incomplete wave data of the sources $F_i$. The high dimensional weight vector of the UBP formula should minimize some error function 
\begin{align}\label{eq:err}
E(\mathbb{W},(G_i,F_i)_{i=1}^M)=\frac{1}{M}\sum_{i=1}^M\mathcal{D}(F_i,P(\mathbb{W},G_i)), 
\end{align}
where $\mathcal{D}:\R^{N_x\times N_x}\times \R^{N_x\times N_x}\rightarrow [0,\infty)$ is some distance measure.
The implementation of the network was done in keras, running on top of tensorflow, where a first layer of the network does the filtering in time without any trainable weights and a second layer implements the backprojection with the adjustable weights $\mathbb{W}$.

\section{Results and discussion}
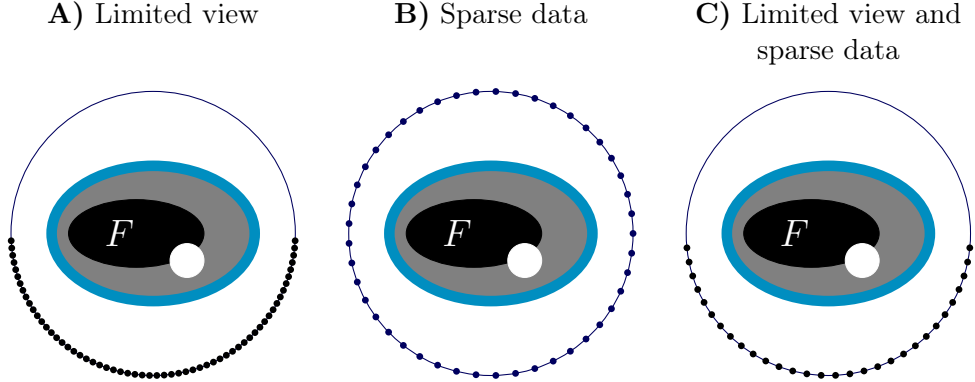
\begin{figure}[ht!!]
\begin{minipage}[t]{0.29\linewidth}
\begin{tikzpicture}[scale=0.45]
\draw (0,7) node{};
\draw (0,-5.3) node{};
\filldraw[color=gray] (0,0) ellipse (3cm and 2cm);
\draw[line width=4pt,color=lime] (0,0) ellipse (3cm and 2cm);
\filldraw[color=black] (-0.5,0) ellipse (2cm and 1cm);
\draw[color=white] (-1,0) node{{\Large $F$}};
\filldraw[color=white] (1,-0.8) circle(0.5);
\draw [color=dot](0,0) circle(4.2);
\equic[4.2 cm]{60}{0}
\draw (0,5.7) node[above] {\textbf{A)}  Limited view};
\end{tikzpicture}
\end{minipage}%
\hfill
\begin{minipage}[t]{0.29\linewidth}
\begin{tikzpicture}[scale=0.45]
\draw (0,7) node{};
\draw (0,-5.3) node{};
\filldraw[color=gray] (0,0) ellipse (3cm and 2cm);
\draw[line width=4pt,color=lime] (0,0) ellipse (3cm and 2cm);
\filldraw[color=black] (-0.5,0) ellipse (2cm and 1cm);
\draw[color=white] (-1,0) node{{\Large $F$}};
\filldraw[color=white] (1,-0.8) circle(0.5);
\draw [color=dot](0,0) circle(4.2);
\equicfull[4.2 cm]{45}{0}
\draw (0,5.7) node[above] {\textbf{B)}  Sparse data};
\end{tikzpicture}
\end{minipage}%
\hfill
\begin{minipage}[t]{0.29\linewidth}
\begin{tikzpicture}[scale=0.45]
\draw (0,7) node{};
\draw (0,-5.3) node{};
\filldraw[color=gray] (0,0) ellipse (3cm and 2cm);
\draw[line width=4pt,color=lime] (0,0) ellipse (3cm and 2cm);
\filldraw[color=black] (-0.5,0) ellipse (2cm and 1cm);
\draw[color=white] (-1,0) node{{\Large $F$}};
\filldraw[color=white] (1,-0.8) circle(0.5);
\draw [color=dot](0,0) circle(4.2);
\equic[4.2 cm]{30}{0}
\draw (0,5.7) node[above] {\textbf{C)}  Limited view and};
\draw (0,4.7) node[above] {sparse data};
\end{tikzpicture}
\end{minipage}
\caption{Considered measurement scenarios. The detectors $\{s_k\}_{k=1}^{N_s}\subset \Gamma$ are represented by dots and $\partial\Omega$ by the solid line.}\label{fig:meas}
\end{figure}

In our numerical studies we consider the case where $\Omega=B_1$ is the ball around the origin with radius one.
We learn weights in the UBP for three different limited data scenarios, shown in Figure \ref{fig:meas}. To efficiently optimize the weights we use the deep learning framework keras \cite{chollet2015keras} which is based on tensorflow \cite{abadi2016tensorflow}. Using this efficient implementation the weighted UBP formula can be combined with CNNs to further improve reconstructions.\\

For learning the weights in all three cases, we numerically compute direction dependent wave data \eqref{eq:dirsol} for $M=1800$ data-source pairs, with discretized sources $F_i\in\R^{256\times 256}$. These generated initial sources are variations of a Shepp-Logan phantom. In particular the positions and amplitudes of the ellipses in the Shepp-Logan are randomly selected. Additionally, the phantoms were rotated by a randomly chosen angle and some finer structures were added. Finally we performed elastic deformations on the phantoms generated this way. The time discretization for all three scenarios is given by 400 equidistant samples in the interval $[0,3]$. For the error function we choose the distance measure induced by $\|\cdot\|_2$, which gives the mean squared error 
\begin{align}
E(\mathbb{W},(G_i,F_i)_{i=1}^M)=\frac{1}{M}\sum_{i=1}^M\|F_i-P(\mathbb{W},G_i)\|_2.
\end{align} 
The network incorporates about two million free parameters and the minimization is done by the stochastic gradient descent algorithm with batch size one over 100 epochs.

\subsection{Limited view}
\begin{figure}[h!!]%
	\centering
	\subfloat{\includegraphics[trim=0 1.5cm 0 0.8cm,clip,width=0.33\textwidth]{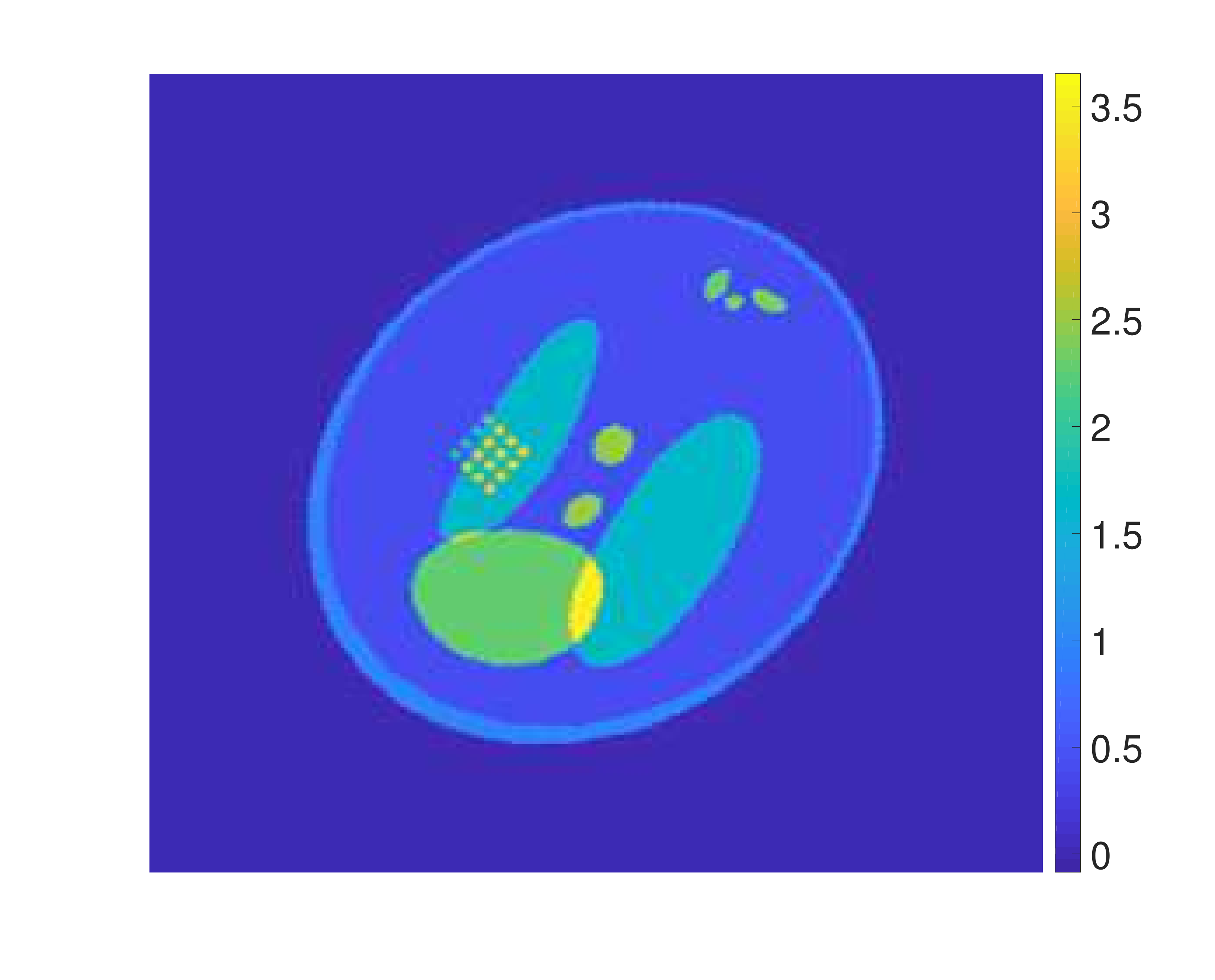}}%
	\subfloat{\includegraphics[trim=0 1.5cm 0 2cm,clip,width=0.33\textwidth]{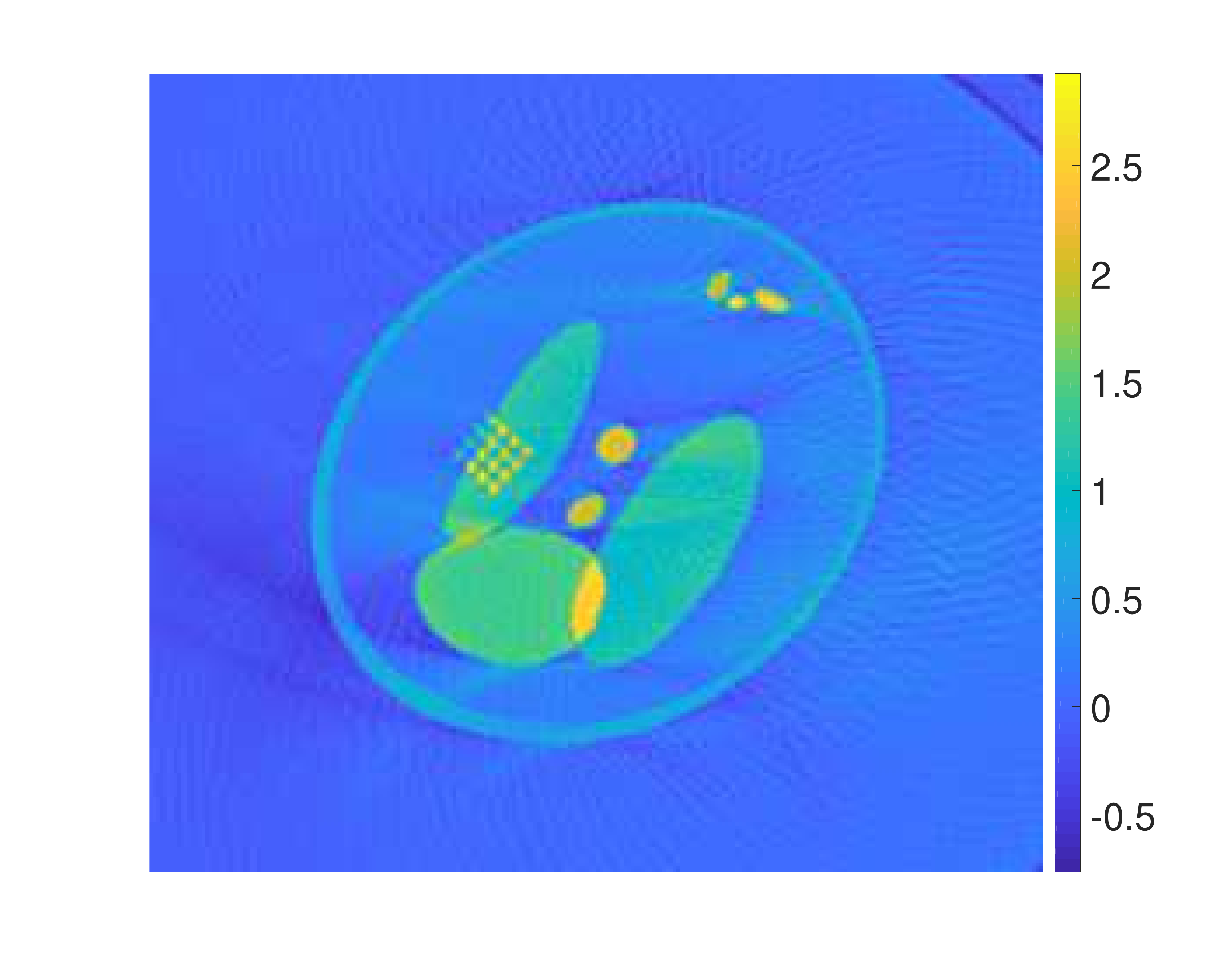}}%
	\subfloat{\includegraphics[trim=0 1.5cm 0 2cm,clip,width=0.33\textwidth]{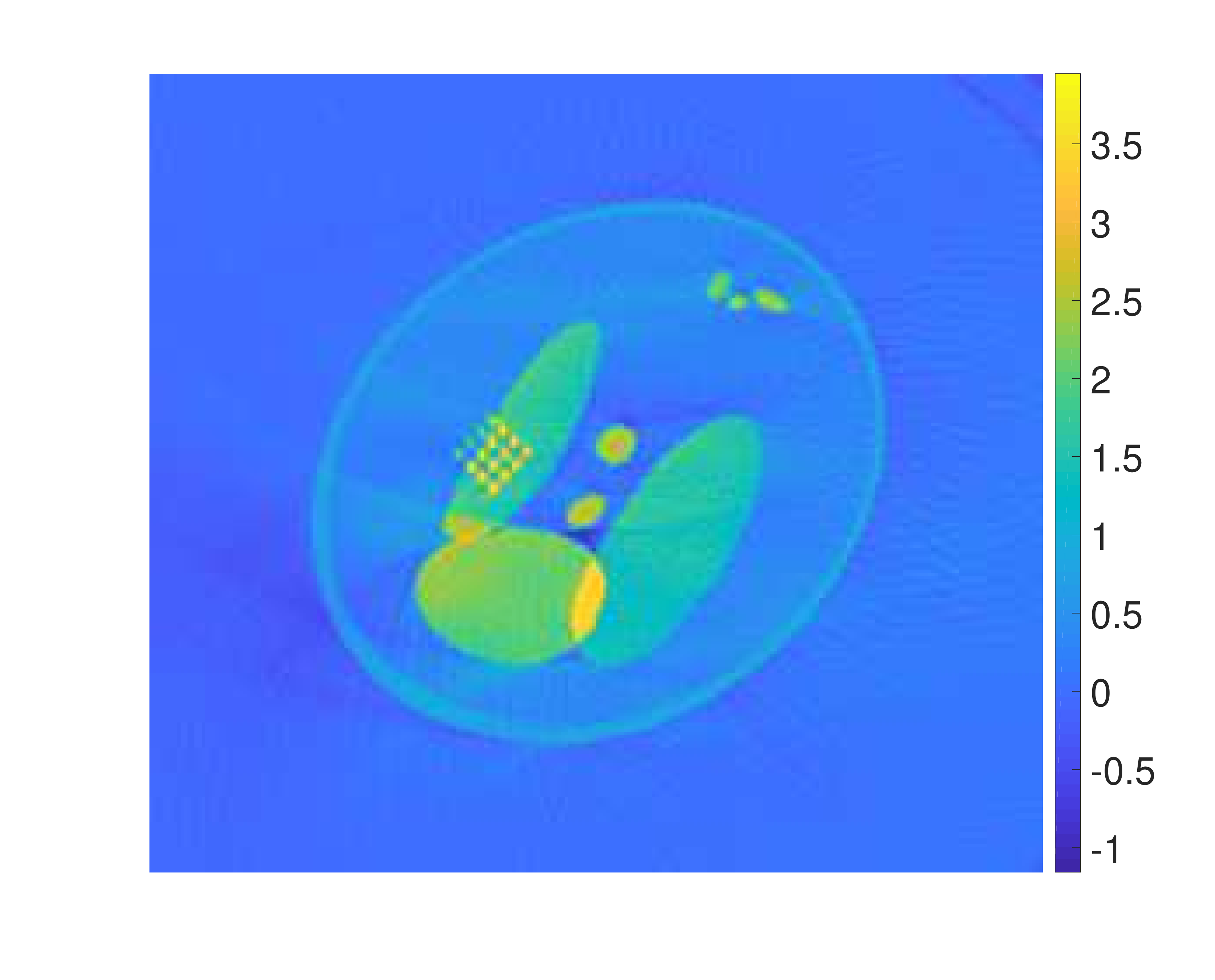}}\\
	\subfloat{\includegraphics[trim=0 1.5cm 0 0.8cm,clip,width=0.33\textwidth]{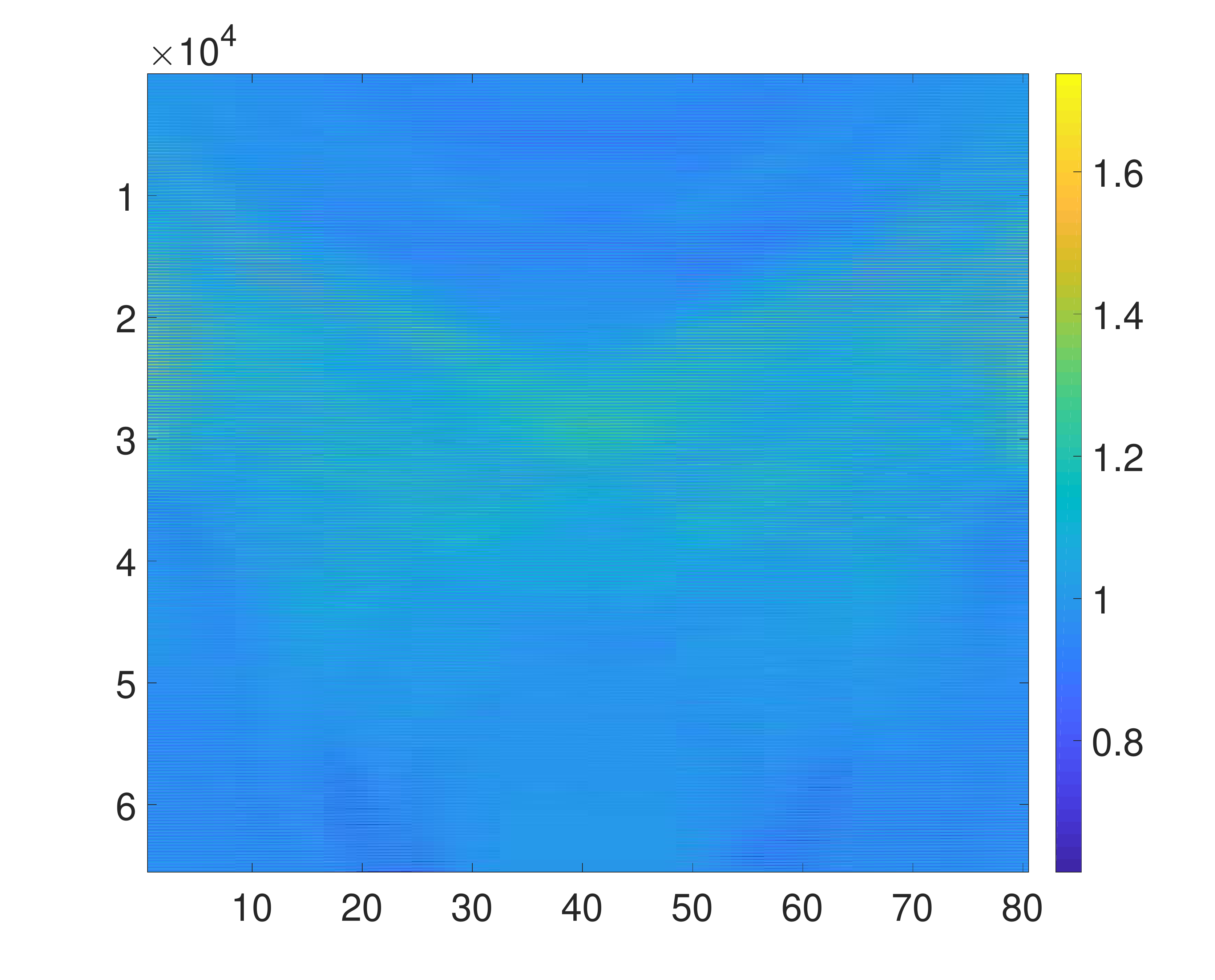}}%
	\subfloat{\includegraphics[trim=0 1.5cm 0 2cm,clip,width=0.33\textwidth]{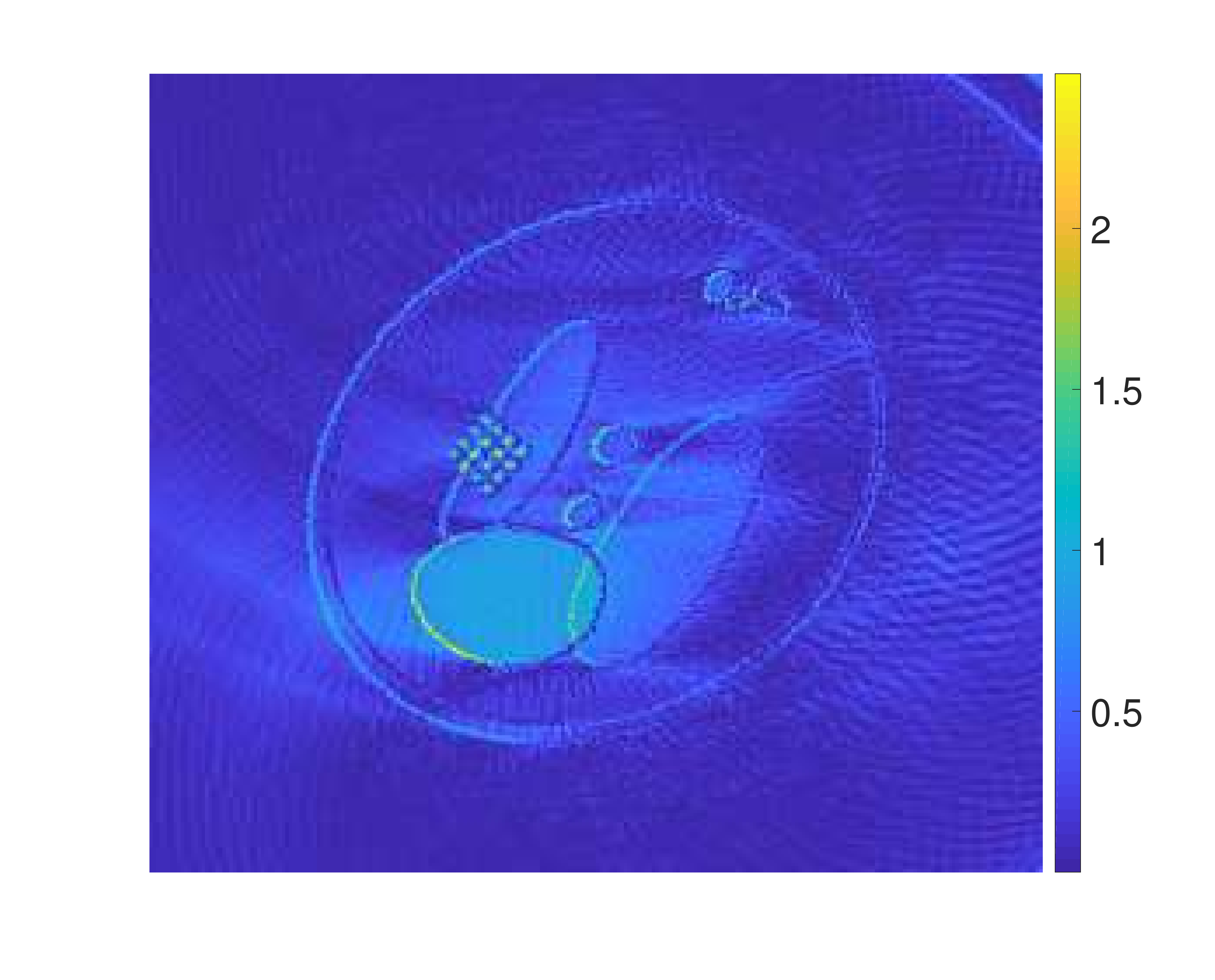}}%
	\subfloat{\includegraphics[trim=0 1.5cm 0 2cm,clip,width=0.33\textwidth]{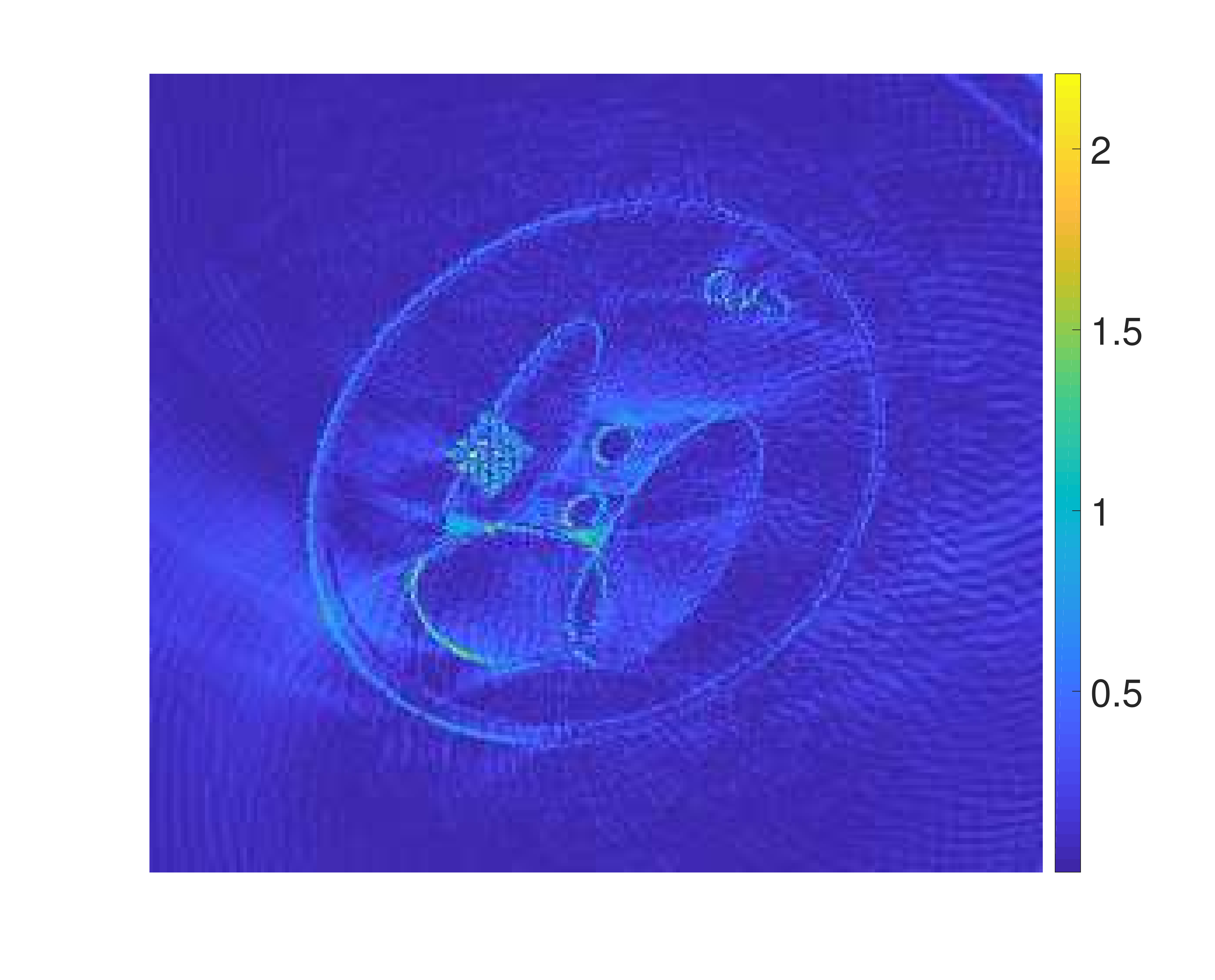}}\\
	\caption{ \textbf{Reconstruction results for the limited view case in the reconstruction domain} $[-1,1]^2$. In the top row a simulated phantom not contained in the training set (left), the reconstruction using the UBP without trained weights (center) and the reconstruction using the weighted UBP (right) are shown. The bottom row displays the learned weights for measurements on a half circle (left) and the absolute differences from the plain UBP (center) and weighted UBP (right) to the ground truth.}\label{fig:train1}
\end{figure}

For the limited view case we assumed the detectors to lie on the half circle as shown in Figure \ref{fig:meas} \textbf{A)}. The simulated photoacoustic data was computed on 100 detector locations. The results for a phantom randomly selected according to the model described above but not contained in the training set is shown in Figure \ref{fig:train1}.

As can be seen in Figure \ref{fig:train1} the reconstructions with the weighted UBP leads slightly better results than the reconstruction with the unmodified UBP. We also observe, that the learned weights follow a certain structure depending on the angle between the reconstruction and detection point.

\subsection{Sparse sampling}

\begin{figure}[h!!]%
	\centering
	\subfloat{\includegraphics[trim=0 1.5cm 0 0.8cm,clip,width=0.33\textwidth]{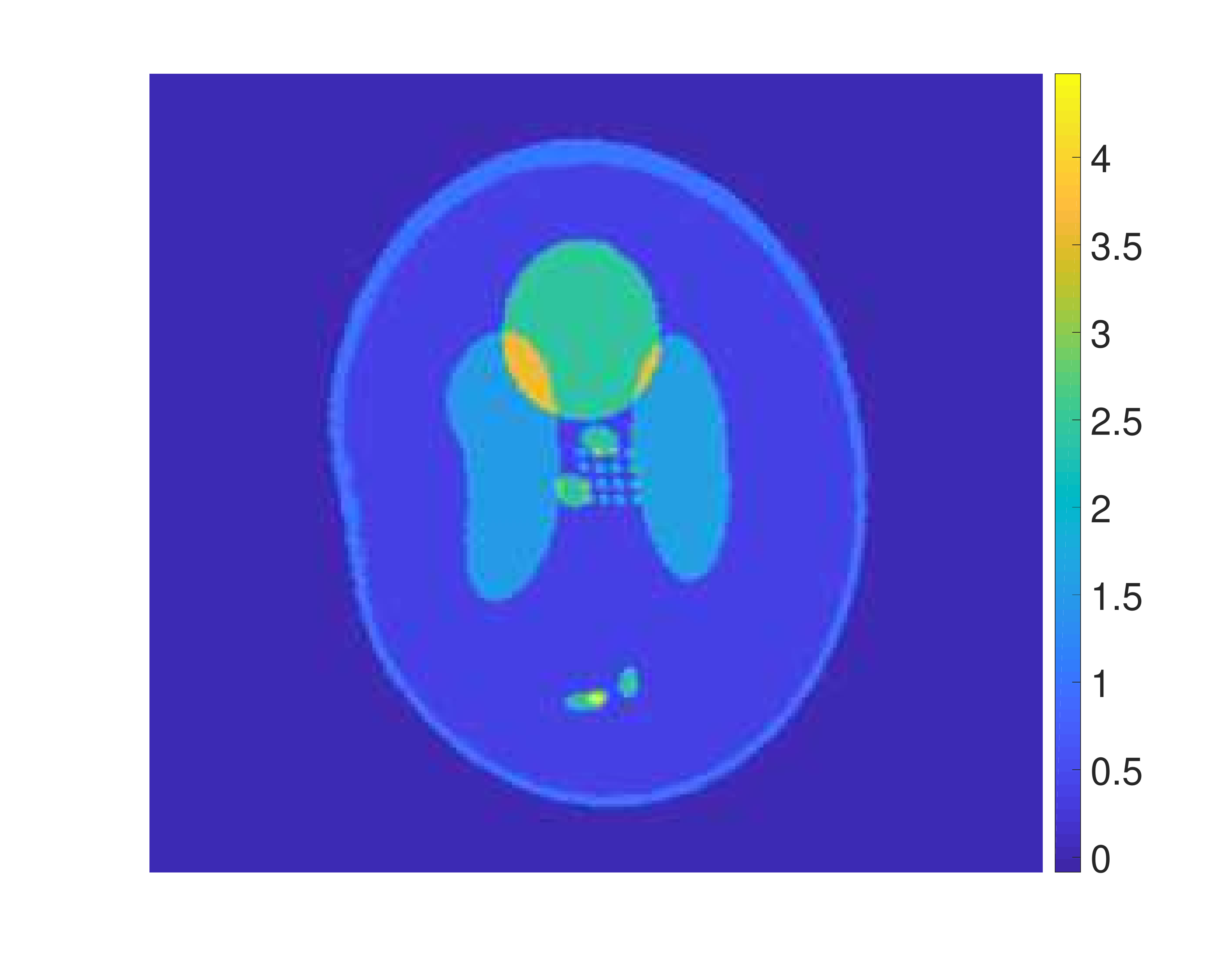}}%
	\subfloat{\includegraphics[trim=0 1.5cm 0 2cm,clip,width=0.33\textwidth]{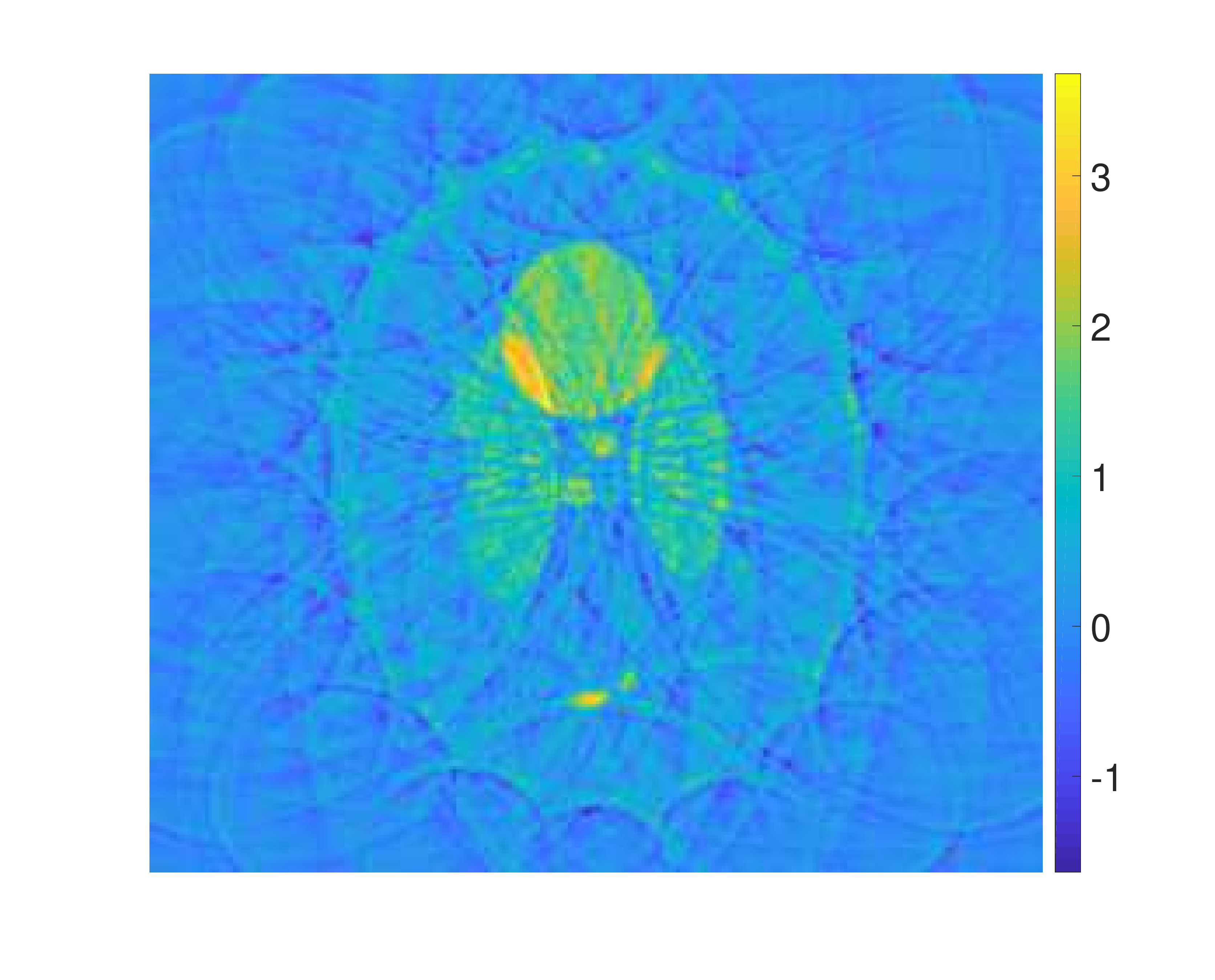}}%
	\subfloat{\includegraphics[trim=0 1.5cm 0 2cm,clip,width=0.33\textwidth]{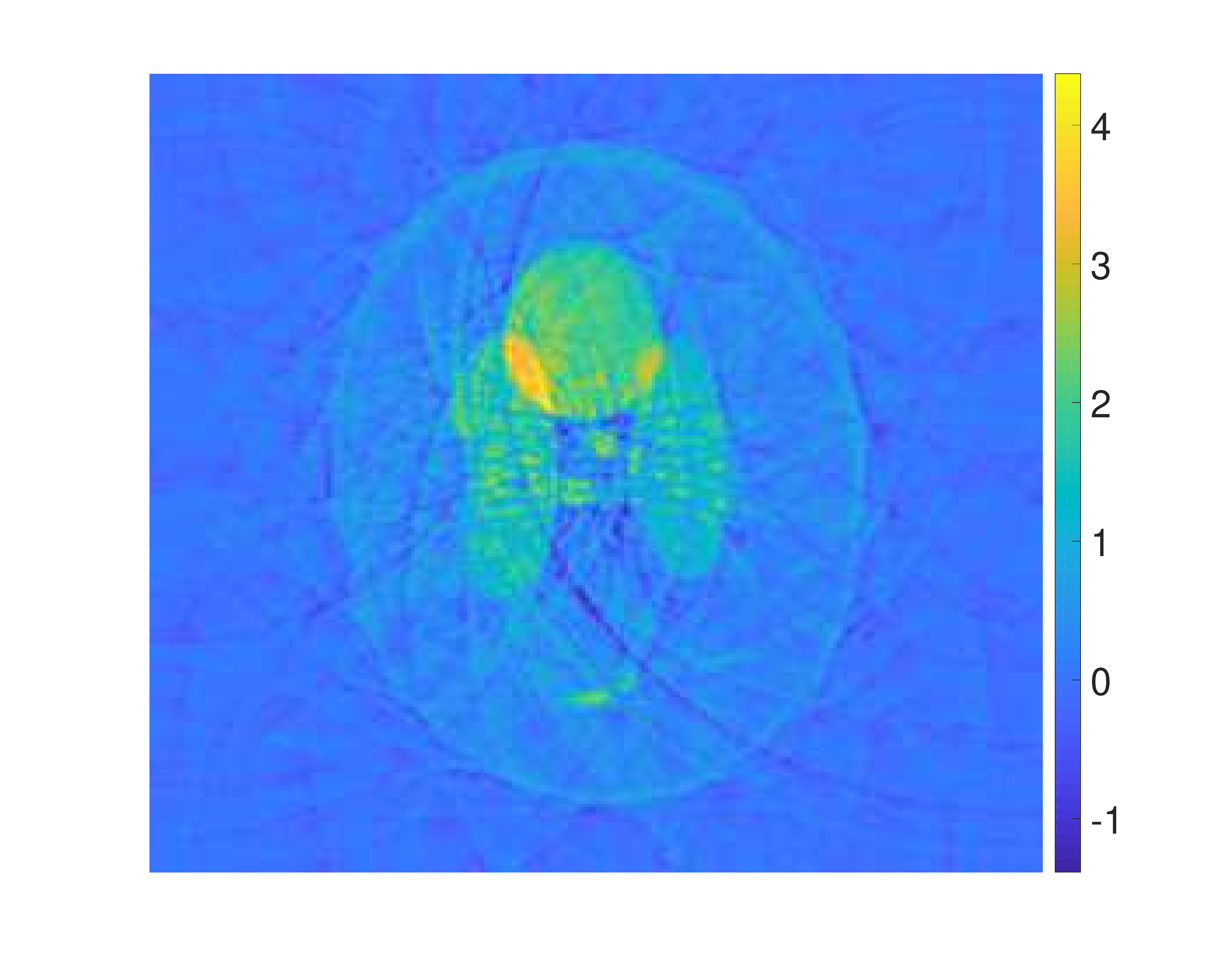}}\\
	\subfloat{\includegraphics[trim=0 1.5cm 0 0.8cm,clip,width=0.33\textwidth]{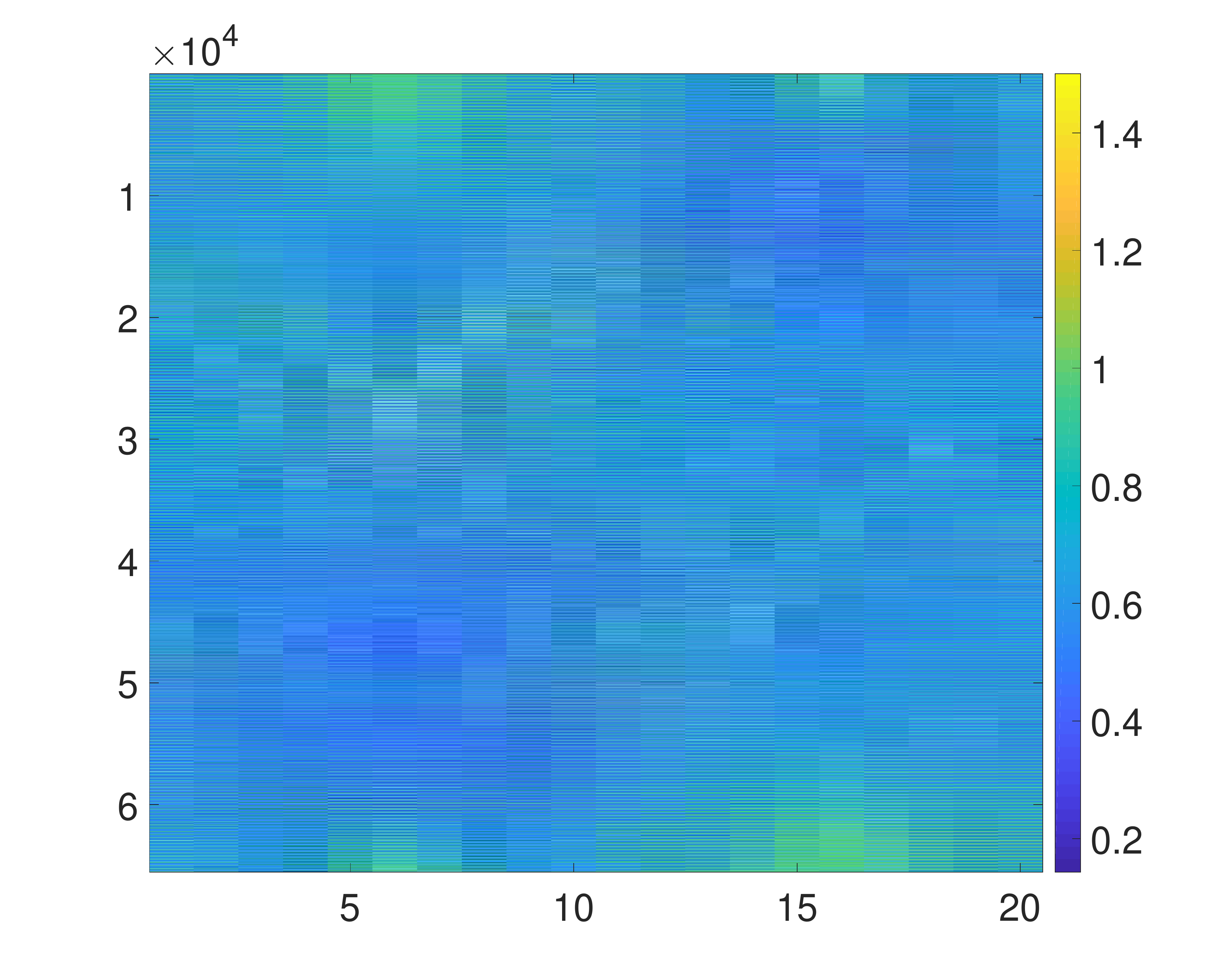}}%
	\subfloat{\includegraphics[trim=0 1.5cm 0 2cm,clip,width=0.33\textwidth]{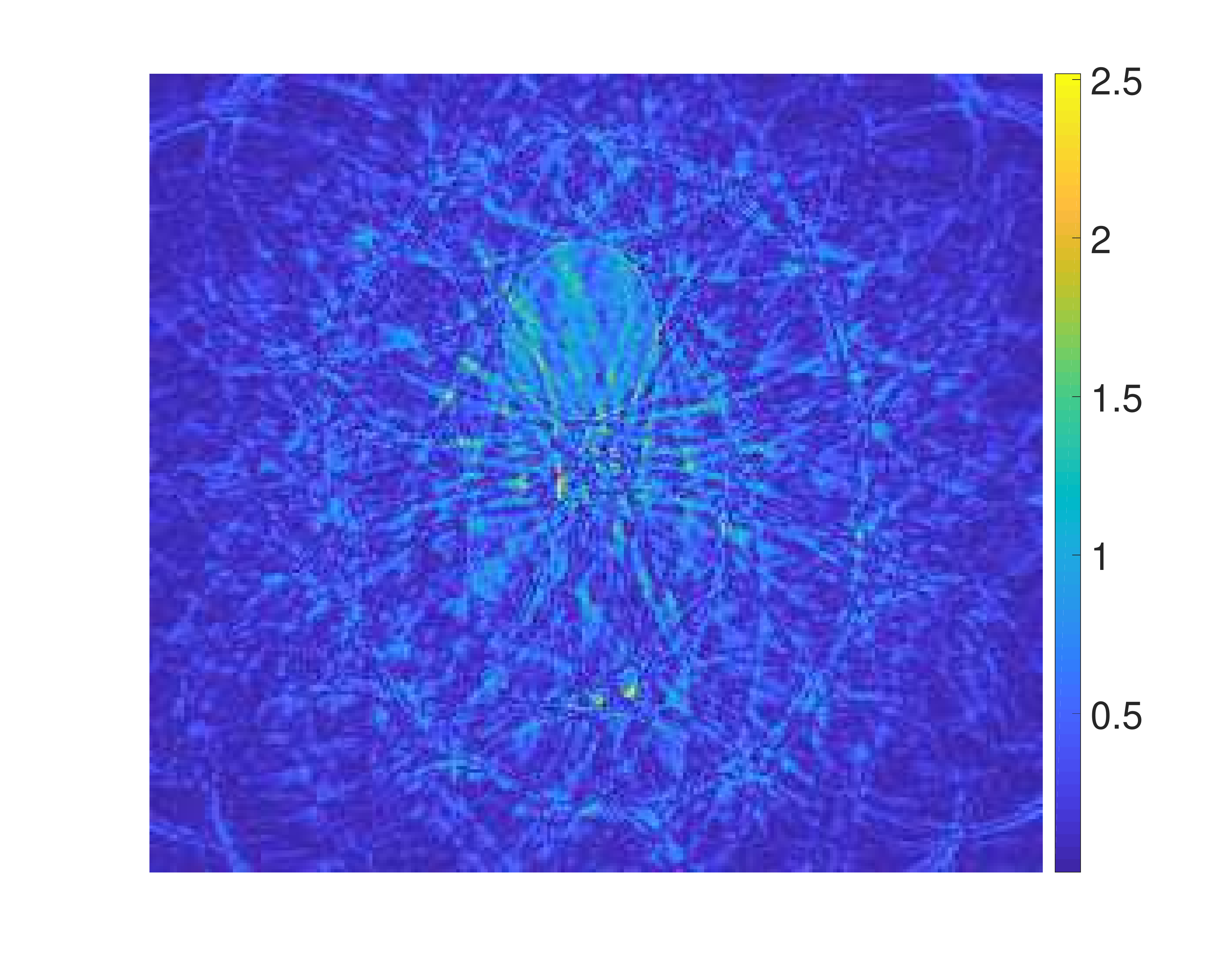}}%
	\subfloat{\includegraphics[trim=0 1.5cm 0 2cm,clip,width=0.33\textwidth]{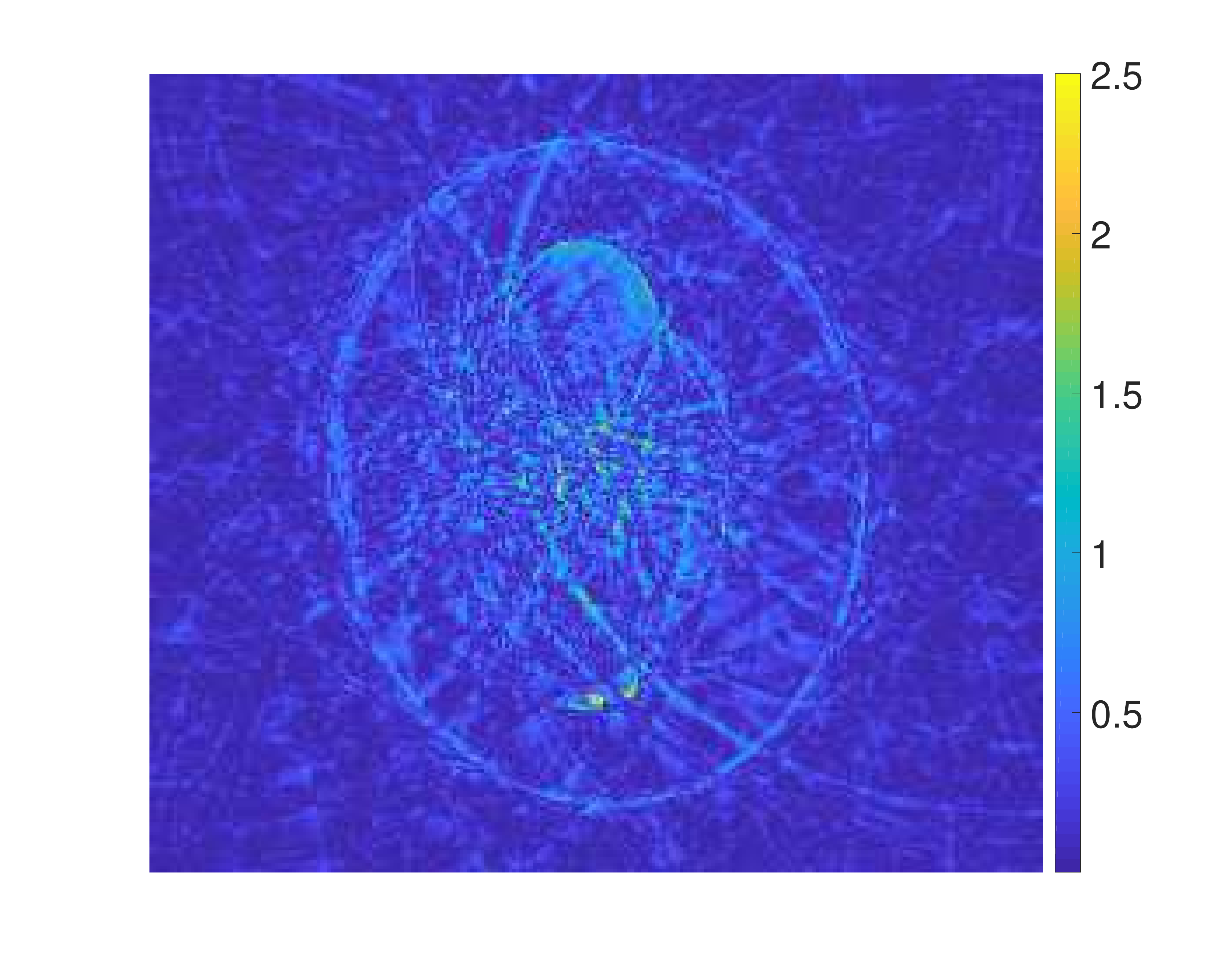}}\\
	\caption{ \textbf{Reconstruction results for sparse data on the reconstruction domain} $[-1,1]^2$. In the top row a simulated phantom not contained in the training set (left), the reconstruction using the UBP without trained weights (center) and the reconstruction using the weighted UBP (right) are shown. The third row displays the learned weights for sparse measurements and the absolute differences from the plain UBP (center) and weighted UBP (right) to the ground truth.}\label{fig:train2}
\end{figure}

In the sparse data scenario the measurement geometry was taken to be a circle with $N_s=20$ equidistant measurement positions, see Figure \ref{fig:meas} \textbf{B)}. Reconstruction results in the sparse data case are presented in Figure \ref{fig:train2}.

Again, the reconstructions using the learned UBP are visually superior to the results obtained with the unweighted UBP formula.

\subsection{Limited view and sparse sampling}

\begin{figure}[h!!]%
	\centering
	\subfloat{\includegraphics[trim=0 1.5cm 0 0.8cm,clip,width=0.33\textwidth]{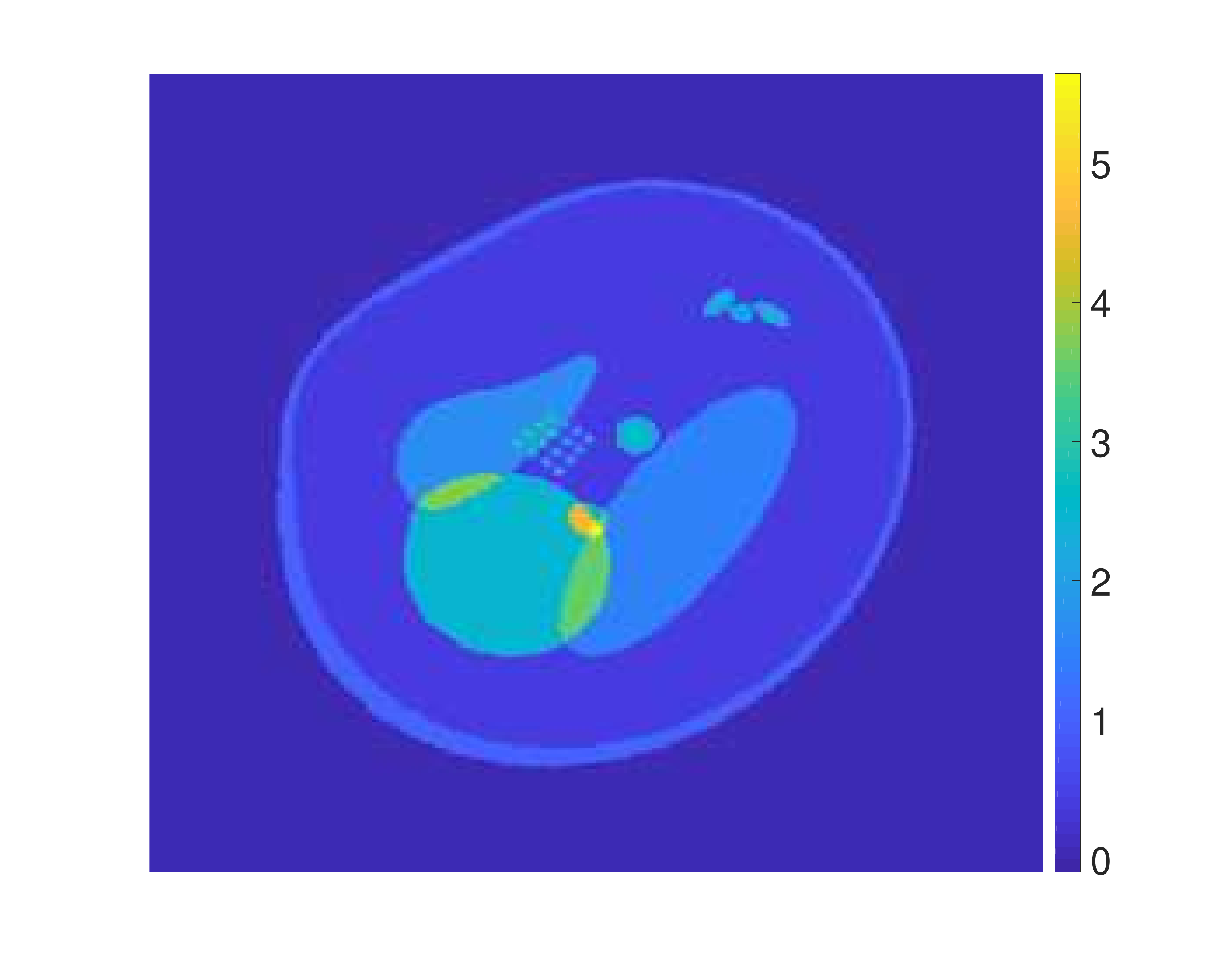}}%
	\subfloat{\includegraphics[trim=0 1.5cm 0 2cm,clip,width=0.33\textwidth]{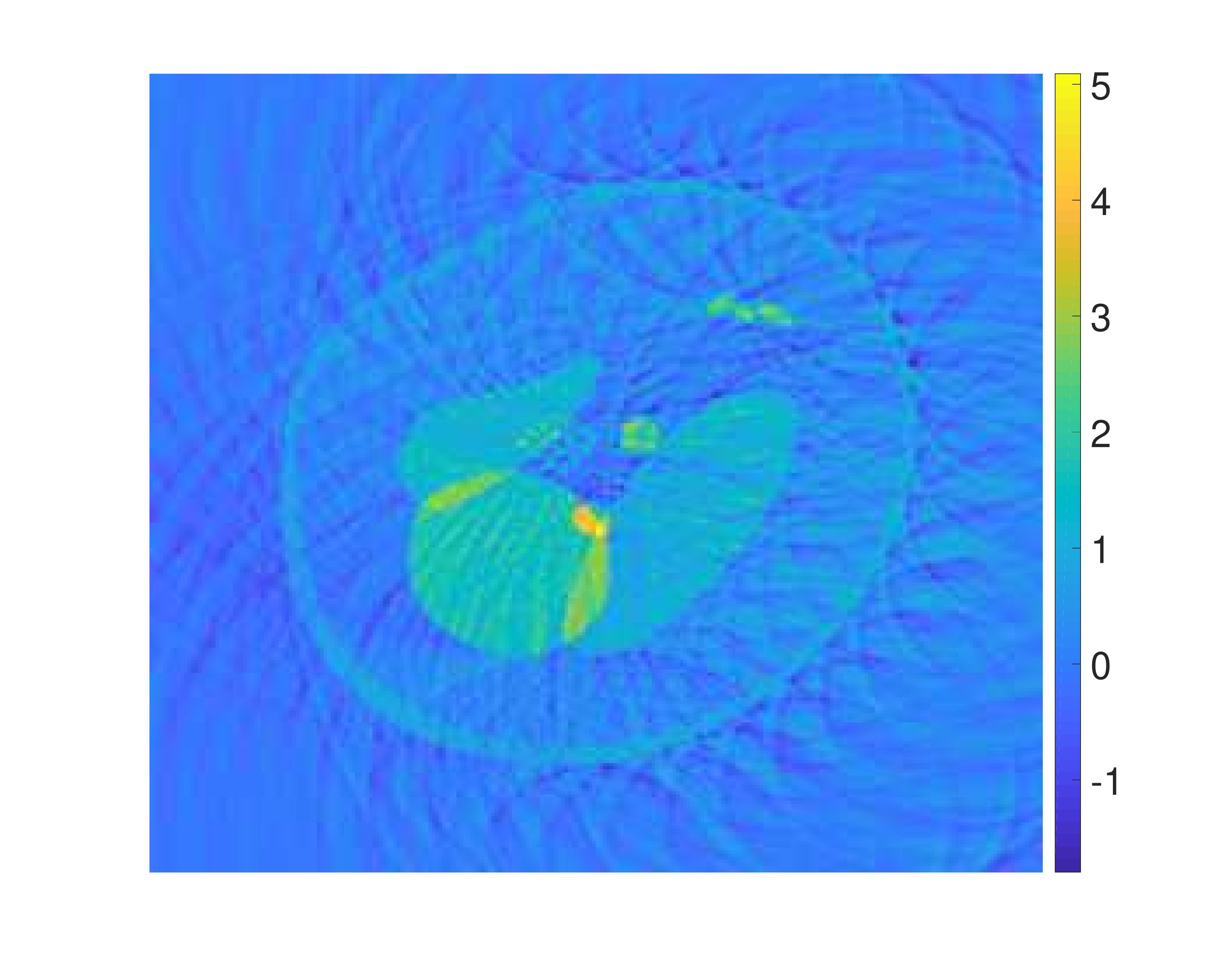}}%
	\subfloat{\includegraphics[trim=0 1.5cm 0 2cm,clip,width=0.33\textwidth]{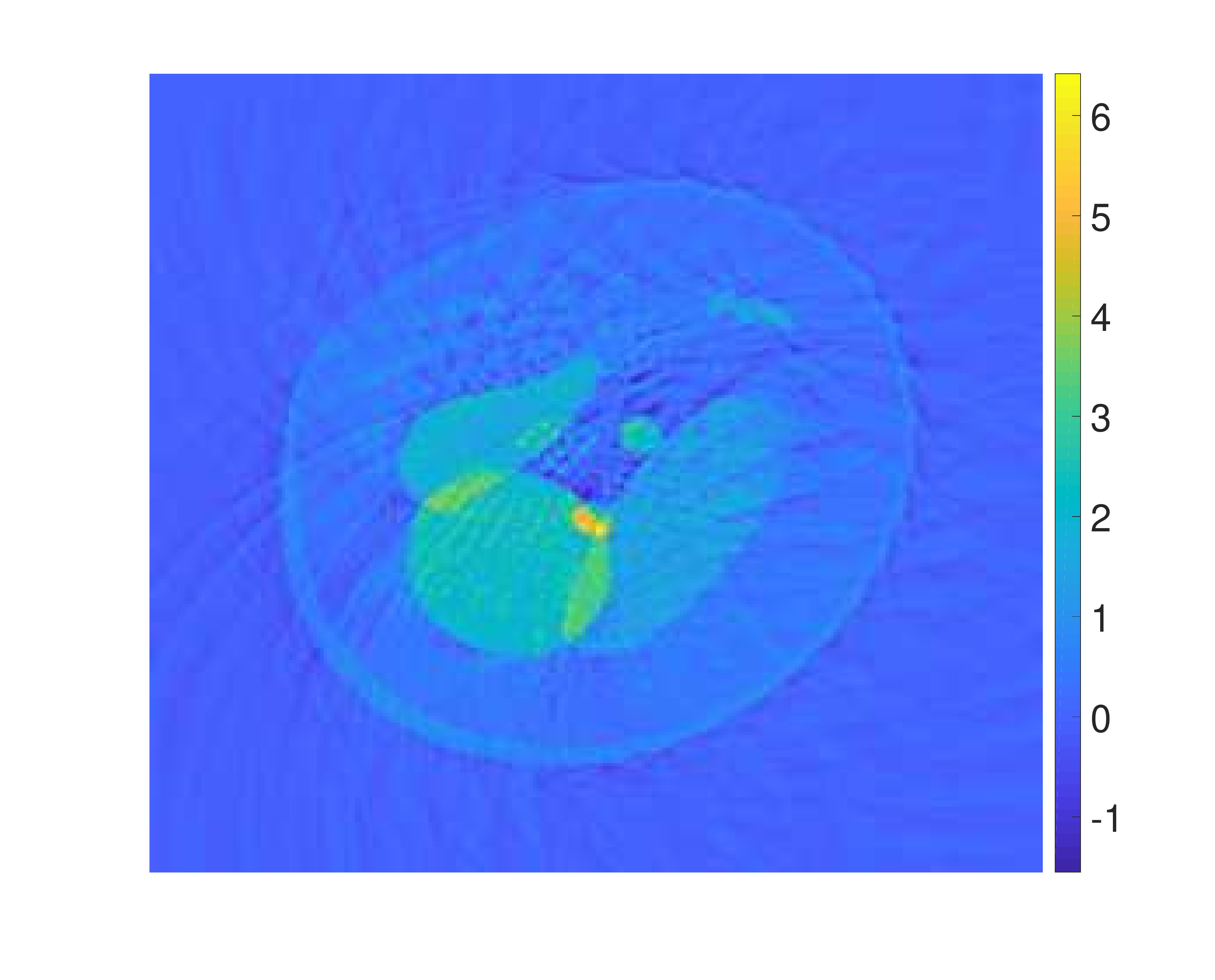}}\\
	\subfloat{\includegraphics[trim=0 1.5cm 0 0.8cm,clip,width=0.33\textwidth]{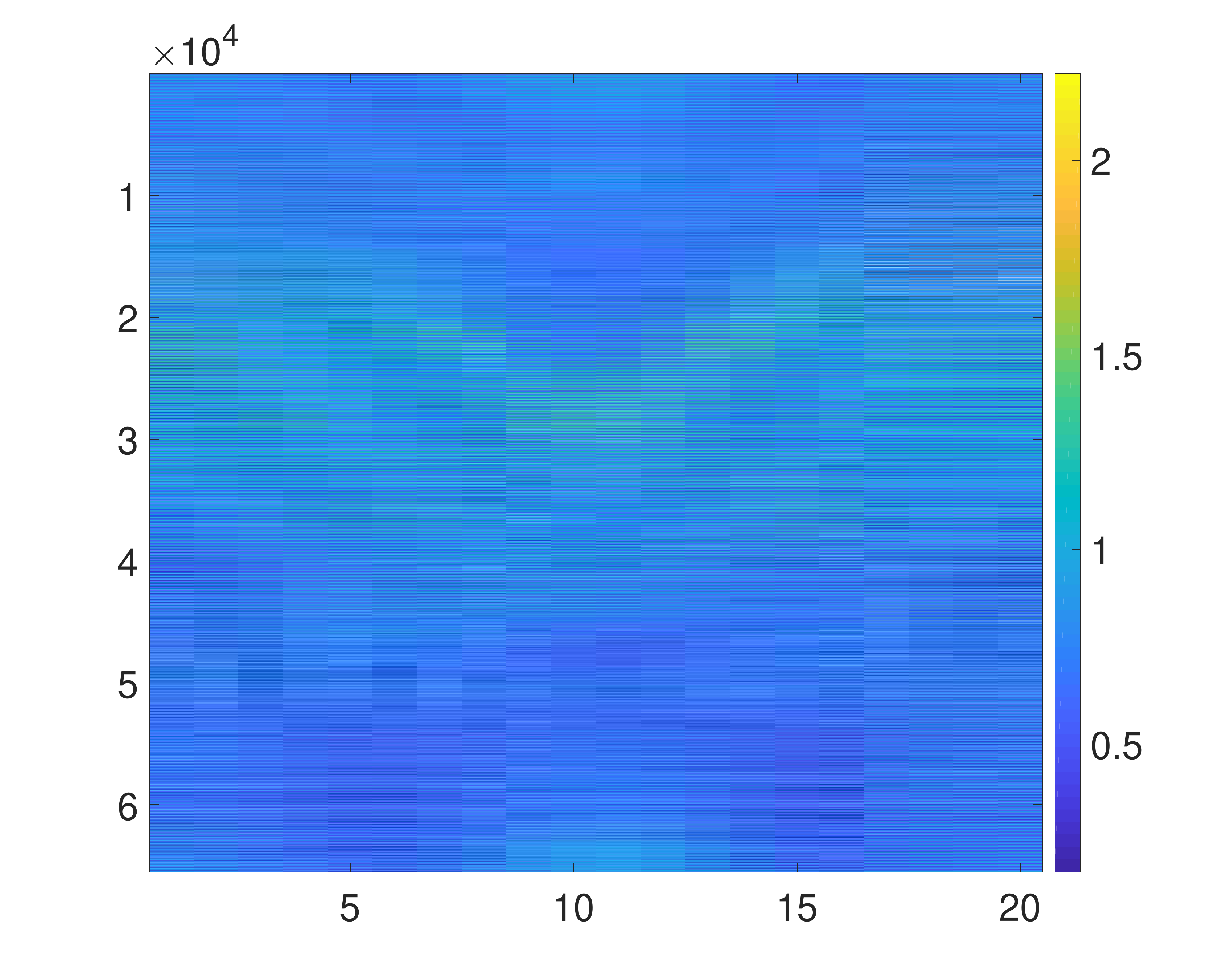}}%
	\subfloat{\includegraphics[trim=0 1.5cm 0 2cm,clip,width=0.33\textwidth]{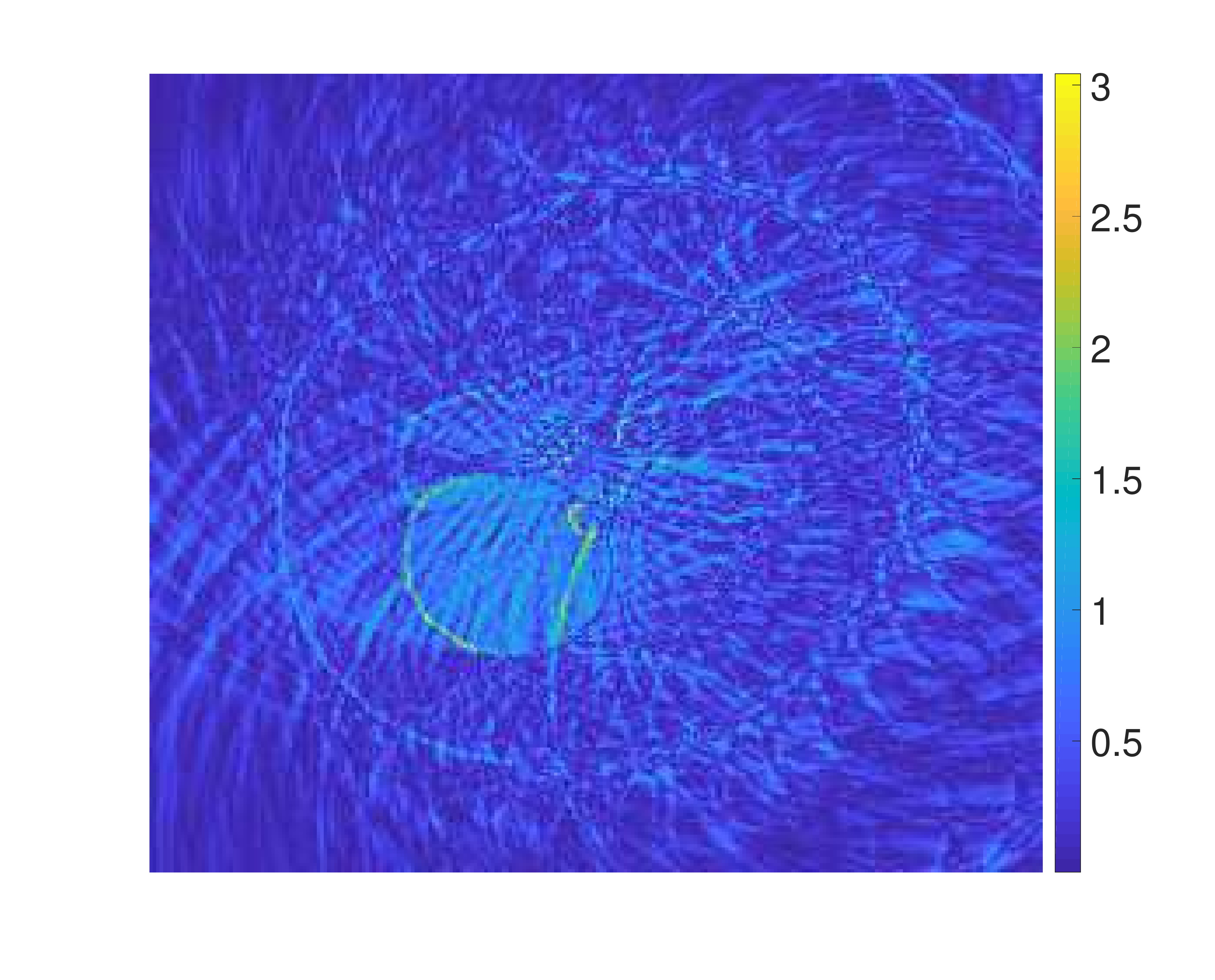}}%
	\subfloat{\includegraphics[trim=0 1.5cm 0 2cm,clip,width=0.33\textwidth]{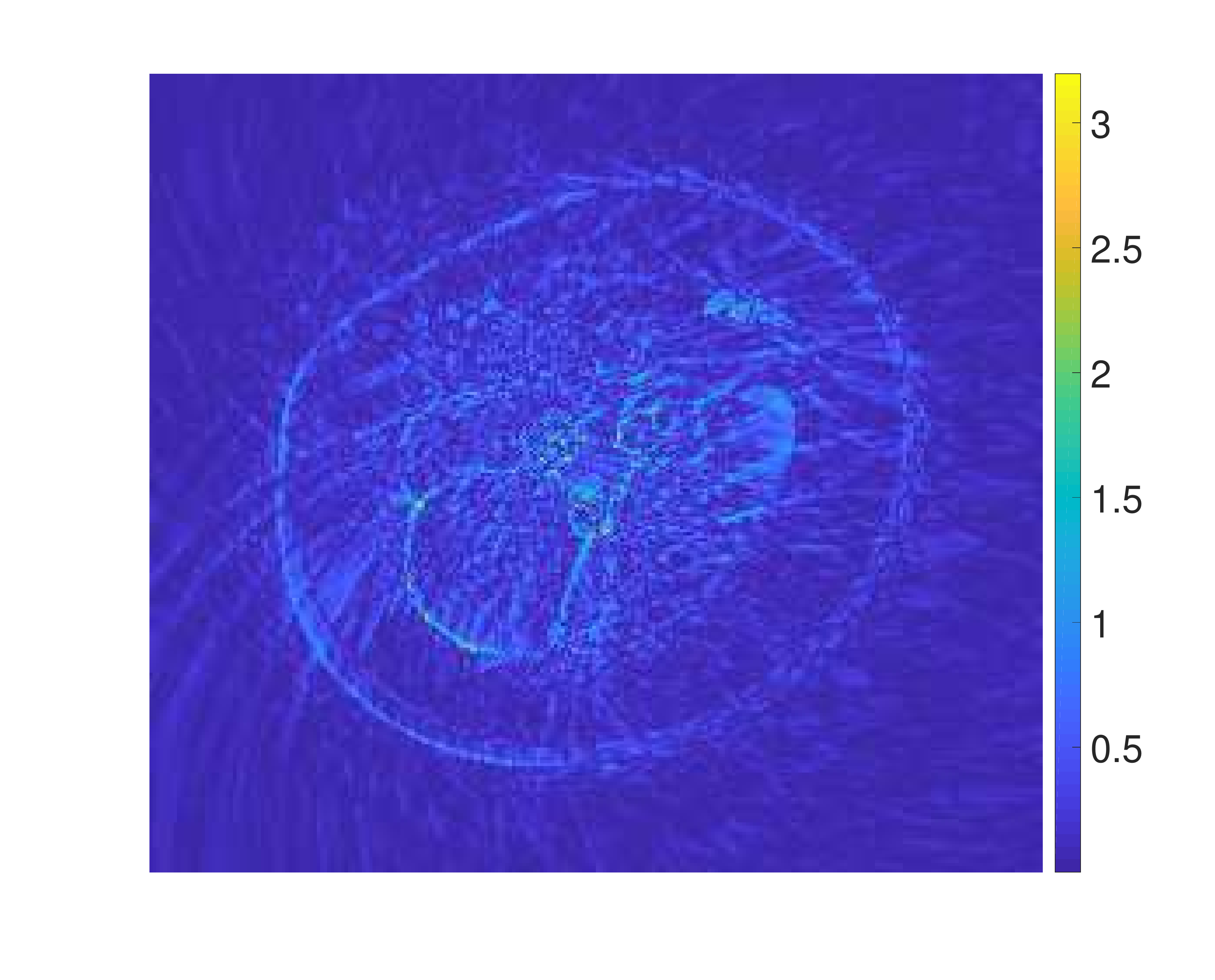}}\\
	\caption{\textbf{Reconstruction results for limited view and sparse data on the reconstruction domain} $[-1,1]^2$. In the top row a simulated phantom not contained in the training set, the reconstruction using the UBP without trained weights and the reconstruction using the weighted UBP are shown. The third row shows the learned weights for sparse measurements on the half circle and the absolute differences to the ground truth.}\label{fig:train3}
\end{figure}

In the limited view and sparse sampling case the detector locations were chosen to lie on a half circle as shown in \ref{fig:meas} \textbf{C)}. The wave data was computed on $N_s=20$ detector positions. The reconstruction results for these case are displayed in Figure \ref{fig:train3} and are once more superior to the reconstructions with the ordinary UBP. 

\subsection{Quantitative evaluation}

To evaluate of the learned universal backprojection formula not just visually the relative squared $\ell^2$-error averaged over 200 reconstructions $(\hat{F_i})_{i=1}^{200}$ of phantoms $(F_i)_{i=1}^{200}$ not contained in the training set
\begin{equation*}
\operatorname{error}\left((F_i)_{i=1}^{200},(\hat{F_i})_{i=1}^{200}\right)\coloneqq \frac{1}{200}\sum_{i=1}^{200}\frac{\|\hat{F_i}-F_i\|_2}{\|F_i\|_2},
\end{equation*}
was computed.
The averaged reconstruction errors for all scenarios \textbf{A)}, \textbf{B)}, \textbf{C)} are displayed in the following table.

\begin{table}[h]
\centering
\begin{tabular}{p{4cm}|p{2cm}|p{2cm}|p{2cm}}
&\textbf{A)} &\textbf{B)} &\textbf{C)}\\
\hline
UBP  &0.2002 &0.3461 &0.3546\\
\hline
weighted UBP & 0.0912 &0.1806 &0.1649
\end{tabular}
\caption{Reconstruction errors for the different scenarios \textbf{A)} limited view, \textbf{B)} sparse sampling and \textbf{C)} sparse sampling and limited view.}\label{tab:err}
\end{table}

Table \ref{tab:err} shows that the reconstruction error is roughly halved by the use of adjusted weights in the UBP formula.

\section{Conclusion}
In this paper, we proposed learned weight factors to improve the reconstruction quality of filtered backprojection algorithms in PAT accounting for limited view data, sparse sampling and detector directivity. The optimization procedure of the weight factors 
is very flexible and can be done for arbitrary surfaces, including geometries, where no exact inversion formula is known. One only needs an appropriate training set consisting of PA sources and corresponding PA data. The presented results demonstrate, that for data given on a half circle, the learned FBP clearly improves the results compared the standard FBP. 

We note that the learnable backprojection could also be used as first layer in a deep CNN as proposed in \cite{schwab2018dalnet}. Further, it is possible to improve the reconstruction quality by learning temporal filters in the UBP, or to use correction weights to reduce the error on arbitrary convex and bounded domains.

\section*{Acknowledgement}
The research of S.A. and M.H. has been supported by the Austrian Science Fund (FWF), project P 30747-N32;
the work of R.N. has been supported by the FWF, project P 28032.
We acknowledge the support of NVIDIA Corporation with the donation of the Titan Xp GPU used for this research.

\bibliographystyle{plain}

\end{document}